\documentclass[usenatbib]{mn2e}
\usepackage{amssymb}

\usepackage{graphicx}
 
\title[Orbital Parameters of the Microquasar LS~I~+61~303]
{Orbital Parameters of the Microquasar LS~I~+61~303}

\author[J. Casares et al.]
{J. Casares$^1$, I. Ribas$^2$, J.M. Paredes$^3$, J. Mart\'\i{}$^4$,
C. Allende Prieto$^5$\\
$^1$ Instituto de Astrof\'{\i}sica de Canarias, 38200 La Laguna, Tenerife, 
Spain\\
$^2$ Institut d'Estudis Espacials de Catalunya/CSIC, C/Gran Capit\`a 2-4, Edif.
Nexus, 08034 Barcelona, Spain\\ 
$^3$ Departament d'Astronomia i Meteorologia, Universitat de Barcelona, Av.
Diagonal 647, 08028 Barcelona, Spain\\
$^4$ Departamento de F\'{\i}sica, Escuela Polit\'ecnica Superior, Universidad 
de Ja\'en, Virgen de la Cabeza 2, 23071 Ja\'en, Spain\\
$^5$ McDonald Observatory and Department of Astronomy, University of Texas,
Austin, TX 78712-1083, USA}

\begin{document}

\maketitle

\begin{abstract} 
\noindent New optical spectroscopy of the HMXB microquasar LS~I~+61~303 is
presented.  Eccentric orbital fits to our radial velocity measurements
yield updated orbital parameters in good agreement with previous work. Our
orbital solution indicates that the periastron passage occurs at radio
phase 0.23 and the X-ray/radio outbursts are triggered 2.5--4 days after
the compact star passage. The spectrum of the optical star is consistent
with a B0~V spectral type and contributes $\sim$65 percent of the total
light, the remaining due to emission by a circumstellar disc. 
We also measure the projected rotational velocity to be $v \sin i
\simeq 113$ km s$^{-1}$. 
\end{abstract}

\begin{keywords}
stars: accretion, accretion discs -- binaries:close -- stars: individual
(LS~I~+61~303) -- X-rays: stars
\end{keywords}

\section{Introduction}

LS~I~+61~303 is a peculiar radio and X-ray star which has attracted the
interest of astronomers during the last quarter of a century. Attention to
it was first called by \cite{gregory78} during a galactic plane survey
aimed at the identification of radio variables on both short and long
time-scales. This precursor work unveiled what has remained a key property
of LS~I~+61~303 over the years, that is, a clearly periodic modulation of
its radio emission every 26.5~days \citep{Taylor82}. 

Just to mention the key facts, LS~I~+61~303 is a high mass X-ray binary 
(HMXB) with the optical component being a B0~Ve star with a circumstellar 
disc, $V\simeq 10.7$ mag, and with an estimated distance of 2 kpc
\citep{Hutchings&crampton81,Paredes86,Frail91}. The periodic modulation
has also been observed at optical and infrared wavelengths
\citep{Mendelson89,Paredes94}, in X-rays \citep{Paredes97}, and possibly
in $\gamma$-rays as well \citep{Massi04a}. This modulation is widely
attributed to the orbital period of the binary system and the most
accurate determination ($P=26.4960\pm0.0028$ d) comes from radio
observations \citep{Gregory02b}.  A second modulation, on a longer time
scale of $\sim$4 yr, has also been reported \citep{Paredes87,Gregory89}
and is likely associated with an outward-moving density enhancement or
shell in the equatorial disc of the rapidly rotating Be star
\citep{Gregory02}. The recent detection of relativistic radio jets -- at
milli-arcsecond angular scales and with hints of precession -- by
\cite{Massi04} firmly includes LS~I~+61~303 in the selected class of
galactic microquasars.
The nature of the compact object remains unknown given the poor 
constraints in the system parameters and the lack of classical neutron star
signatures (e.g. X-ray pulses or Type I bursts).  

The seminal paper by \cite{gregory78} already pointed out the possible
connection of the peculiar radio star with the COS B $\gamma$-ray source
CG135+01. The idea of LS~I~+61~303 as a $\gamma$-ray emitter has persisted
over the years. Today, this star is the best counterpart candidate to 3EG
J0241+6103, one of the unidentified high energy ($\geq 100$ MeV) sources
in the 3rd EGRET catalog \citep{Hartman99}. Such association is likely not
to be a unique case. The microquasar LS~5039, probably associated with
another EGRET source, brought the idea that microquasars could be behind
some of the unidentified EGRET sources \citep{Paredes00}. Theoretical
modelling of the $\gamma$-ray emission strongly supports this
interpretation. Inverse Compton scattering of stellar photons by the
relativistic electrons in the jets, can account for the overall properties
of microquasars from radio to $\gamma$-rays \citep{Bosch04}.
  
In this context, it is surprising that the orbital parameters of
LS~I~+61~303 have remained poorly known over the years. Over two decades
ago, \cite{Hutchings&crampton81} carried out a series of optical
spectroscopic observations from which a radial velocity curve of the
binary system was determined.  
Spectroscopic determination of the radial velocity curve of LSI+61 303 is
complicated by contamination of the narrow photospheric lines by emission from
the Be circumstellar envelope.
Here we present new optical spectroscopic
observations of LS~I~+61~303 using modern telescopes and detectors. The
idea was to obtain an improved set of spectroscopic orbital elements which
could be used to address issues such as the system's mass function,
orbital eccentricity, or the phase of periastron, among others. An
accurate knowledge of such fundamental properties is esential to remove
the lingering uncertainties that plague all theoretical models for the
multi-wavelength variable emission of LS~I~+61~303, which we believe is a
prime target to attain a better understanding of the $\gamma$-ray sky.

In the following sections we present our observation and reduction
procedures, orbital fits and discussion of results.

\section{Observation and Data Reduction}

LS~I~+61~303 was observed using three different telescopes and instruments
in the blue spectral range ($\lambda\lambda$3400-5800) during years 2002
and 2003. Seventeen intermediate resolution spectra (0.63 \AA\ pix$^{-1}$)
were obtained using the Intermediate Dispersion Spectrograph (IDS)
attached to the 2.5~m Isaac Newton Telescope (INT) at the Observatorio del
Roque de Los Muchachos on the nights of 27-31 July 2002, 13 March 2003 and
1-7 July 2003. We used the R900V grating in combination with the 235~mm
camera and a 1.2$\arcsec$ slit to provide a spectral resolution of 83 km
s$^{-1}$ (FWHM). The spectral type standard $\tau$ Sco was also observed
with the same instrumentation for the purpose of radial velocity analysis.
In addition, this star has a very low projected rotational velocity, lower
than 5 km~s$^{-1}$ according to \citet{Hardorp70} and \citet{Stickland95}.
Six high-resolution spectra of LS~I~+61 303 were also obtained with the
2.7 m Harlan J. Smith telescope and the 2dcoude spectrograph
\citep{Tull95} at McDonald Observatory (Texas) on UT 3-5 August 2002 and
8-9 April 2003. We used the F3 focal station, the E2 echelle grating, and
a 1\farcs2 slit to achieve a FWHM resolution of about 5 km s$^{-1}$.  
Finally, five low resolution spectra (2.74 \AA\ pix$^{-1}$) were obtained
on the nights of 13 October and 15-17 December 2003 with the ALFOSC
spectrograph at the Nordic Optical Telescope (NOT) at the Observatorio del
Roque de Los Muchachos. We employed grism 6 and a 0.5$\arcsec$ slit which
resulted in a spectral resolution of 235 km s$^{-1}$, as measured from
gaussian fits to the arc lines.  A complete log of the observations is
presented in Table \ref{tabobs}.

The images were bias corrected and flat-fielded, and the spectra
subsequently extracted using conventional optimal extraction techniques in
order to optimize the signal-to-noise ratio of the output \citep{Horne86}.  
Frequent observations of comparison arc lamp or hollow cathode lamp images
were performed in the course of each run and the pixel-to-wavelength scale
was derived through polynomial fits to a large number of identified
reference lines. The final rms scatter of the fit was always $<$1/30 of
the spectral dispersion.

\begin{table}
\centering
\caption[]{Log of the observations.}
\label{tabobs}
\scriptsize
\begin{tabular}{lcccc}
\hline
\hline
Date    & Object & Obs. & Exp. Time & Dispersion  \\
 & & & (seconds) &  (\AA\ pix$^{-1}$) \\
\hline
2002-07-27 & LS~I~+61~303& INT & 600,300& 0.63 \\
2002-07-28 & LS~I~+61~303& INT &  2x600 & 0.63 \\
2002-07-29 & LS~I~+61~303& INT & 600,360& 0.63 \\
2002-07-30 & LS~I~+61~303& INT & 2x600  & 0.63 \\
2002-07-31 & LS~I~+61~303& INT &   300  & 0.63 \\
2002-08-03 & LS~I~+61~303& McD & 600    & 0.11 \\
2002-08-04 & LS~I~+61~303& McD &   800  & 0.11 \\
2002-08-05 & LS~I~+61~303& McD &   600  & 0.11 \\
2003-03-13 & LS~I~+61~303& INT & 2x300  & 0.63 \\
2003-04-08 & LS~I~+61~303& McD & 1200   & 0.11 \\
2003-04-09 & LS~I~+61~303& McD & 2x1200 & 0.11 \\
2003-07-01 & LS~I~+61~303& INT &   300  & 0.63 \\
2003-07-01 &$\tau$ Sco   & INT &     1  & 0.63 \\	
2003-07-02 & LS~I~+61~303& INT &   500  & 0.63 \\
2003-07-03 & LS~I~+61~303& INT &   500  & 0.63 \\
2003-07-04 & LS~I~+61~303& INT &   600  & 0.63 \\
2003-07-05 & LS~I~+61~303& INT &   400  & 0.63 \\
2003-07-06 & LS~I~+61~303& INT &   500  & 0.63 \\
2003-07-07 & LS~I~+61~303& INT &   750  & 0.63 \\
2003-12-15 & LS~I~+61~303& NOT &    60  & 2.74 \\
2003-12-16 & LS~I~+61~303& NOT &    60  & 2.74 \\
2003-12-17 & LS~I~+61~303& NOT &    60  & 2.74 \\
\hline
\end{tabular}
\end{table}

\section{Radial Velocities and Orbital Solution}

All the spectra were prepared for the cross-correlation analysis by
subtracting a low order spline fit to the continuum and subsequently
rebinned into a uniform velocity scale of 42 km s$^{-1}$. The template
spectrum, $\tau$ Sco, was broadened by 113 km s$^{-1}$ and 226 km s$^{-1}$
in order to match the width of the photospheric lines in the LS~I~+61~303
spectra obtained with the INT and NOT, respectively.  These broadening
factors were derived by subtracting several broadened versions of $\tau$
Sco from individual spectra of LS~I~+61~303 and performing a $\chi^2$ test
on the residuals \citep*[see details in][]{Marsh94}. To broaden the
template spectrum we employed a Gray rotational profile \citep{Gray92}
with a limb darkening coefficient of $\epsilon=0.33$, appropriate for our
spectral type and wavelength range. Note that the optimal broadening is
driven by differences in instrumental resolution for the case of the NOT
spectra, but not the INT since the same setup was used in both the
LS~I~+61~303 spectra and the template. In the latter case, the broadening
is directly associated to $v \sin i$ and confirms that the optical star in
LS~I~+61~303 is rapidly rotating. On the other hand, since the
instrumental resolution of the McDonald data is much higher than the
template, we decided to optimally broaden the target spectra (to match the
template) for the sake of the cross-correlation analysis. The optimal
broadening was found to be 113 km s$^{-1}$.

We cross-correlated every individual spectrum of LS~I~+61~303 with the
conveniently broadened template of $\tau$ Sco after masking out the
interstellar absorption features at $\lambda$3934 (Ca~{\sc ii} K) and
$\lambda$4430. The correlations include H~{\sc i} lines from H$\beta$ to
H$_8$ and all He~{\sc i} and He~{\sc ii} lines in the spectral range
$\lambda\lambda$3850--5020 (see Fig. 3). 
Furthermore, since the H~{\sc i} lines are
prone to contamination by stellar wind emission in early-type stars, we
also decided to obtain a set of radial velocities with only He~{\sc i} and
He~{\sc ii} lines. The resulting velocities were folded on the radio
ephemeris of \cite{Gregory02b} ($P=26.4960$ d, $T_0$=HJD2443366.775).
Table \ref{tabrvs} lists the radial velocities measured using the two
different sets of spectral lines. 
Individual velocities were extracted following the method of Tonry \& 
Davis (1979), where parabolic fits where performed to the peak of the 
cross-correlation functions, and the uncertainties are purely statistical.

\begin{table}
\centering
\caption[]{Radial velocities of LS~I~+61~303 measured from
cross-correlation.}
\label{tabrvs}
\scriptsize
\begin{tabular}{ccccccc}
\hline
\hline
HJD--    & Phase &\multicolumn{2}{c}{RV (H+He) (km s$^{-1}$)} &
\multicolumn{2}{c}{RV (He) (km s$^{-1}$)} & Obs. \\
2452000  &       &  Value & $\sigma$ & Value & $\sigma$ & \\
\hline
  483.7265 & 0.0878 & $-$48.3 &  2.3 & $-$37.6           &  2.7       & INT\\
  483.7336 & 0.0881 & $-$41.2 &  3.3 & $-$29.5           &  4.1       & INT\\
  484.7173 & 0.1252 & $-$41.1 &  4.5 & $-$34.6           &  5.5       & INT\\
  484.7250 & 0.1255 & $-$44.3 &  2.9 & $-$37.2           &  3.5       & INT\\
  485.6677 & 0.1611 & $-$31.0 &  2.4 & $-$22.1           &  2.9       & INT\\
  485.6735 & 0.1613 & $-$32.2 &  4.9 & $-$20.5           &  6.0       & INT\\
  486.6493 & 0.1982 & $-$22.5 &  2.3 & $-$11.4           &  2.6       & INT\\
  486.6593 & 0.1985 & $-$17.1 &  2.1 & \phantom{1}$-$8.2 &  2.5       & INT\\
  487.6722 & 0.2368 & $-$15.6 &  3.0 & \phantom{1}$-$4.1 &  3.5       & INT\\
  489.9708 & 0.3235 & $-$54.5 &  3.0 & $-$42.2           &  4.2       & McD\\
  490.9739 & 0.3614 & $-$47.4 &  0.8 & $-$42.3           &  1.0       & McD\\
  491.9774 & 0.3992 & $-$56.2 &  1.5 & $-$64.9           &  1.9       & McD\\
  712.3223 & 0.7154 & $-$57.8 &  9.4 & $-$41.8           & \llap{1}1.6& INT\\
  712.3278 & 0.7156 & $-$70.1 &  6.3 & $-$59.0           &  8.0       & INT\\
  737.6144 & 0.6700 & $-$60.0 &  1.1 & $-$40.0           &  1.1       & McD\\
  738.5838 & 0.7066 & $-$49.0 &  1.3 & $-$41.9           &  1.6       & McD\\
  738.6003 & 0.7072 & $-$40.1 &  1.2 & $-$36.7           &  1.4       & McD\\
  822.7259 & 0.8822 & $-$73.1 &  2.0 & $-$64.9           &  2.5       & INT\\
  823.7330 & 0.9202 & $-$58.2 &  1.9 & $-$49.9           &  2.3       & INT\\
  824.7324 & 0.9579 & $-$56.7 &  4.4 & $-$51.2           &  5.4       & INT\\
  825.7301 & 0.9956 & $-$45.2 &  2.4 & $-$34.4           &  2.9       & INT\\
  826.7322 & 0.0334 & $-$52.2 &  2.4 & $-$43.6           &  2.9       & INT\\
  827.7361 & 0.0713 & $-$36.3 &  1.9 & $-$28.7           &  2.3       & INT\\
  828.7309 & 0.1088 & $-$32.0 &  6.9 & $-$26.0           &  8.3       & INT\\
  989.3004 & 0.1690 & $-$31.5 &  6.3 & $-$12.1           &  8.2       & NOT\\
  990.2950 & 0.2065 & $-$29.6 &  4.8 & $-$21.3           &  5.8       & NOT\\
  991.2943 & 0.2442 & $-$38.1 &  4.9 & $-$23.5           &  6.0       & NOT\\
\hline
\end{tabular}
\end{table}

The radial velocity curve was modelled with an eccentric orbital solution.
This was accomplished using the sophisticated binary fitting program
originally developed by \citeauthor{wd71} (\citeyear{wd71}; hereafter
W-D)\footnote{The original code has suffered major upgrades since its
first release, including significant improvements in the underlying
physical models. The most recent version of the W-D program together with
the relevant documentation can be found in {\tt
ftp://astro.ufl.edu/pub/wilson/lcdc2003}.}. Briefly, the W-D program
assumes a Roche model to describe the shape of the components and includes
a complete treatment of the radiative properties of the system (limb
darkening, temperature variations accross the surface, etc.). Thus, the
model adequately takes into account the non-keplerian contributions to the
measured radial velocities. Fitting is carried out through differential
corrections and, in the case of radial velocity solutions, convergence is
achieved very rapidly. In the solutions we adjusted the following orbital
parameters: velocity semiamplitude ($K_1$), systemic velocity ($\gamma$),
eccentricity ($e$), argument of the periastron ($\omega$), and phase of
the periastron ($\phi_{\rm peri}$). A set of fixed parameters necessary to
the fitting procedure were tuned to match the expected radiative
properties of the B0~V optical companion (i.e., $T_{\rm
eff}$$\approx$$28\,000$ K, $\log g$$\approx$$4$; \citeauthor{Cox00}
\citeyear{Cox00}), although the effect of the adopted parameters on the
solutions is very weak. Table \ref{tabrvfit} presents the final
best-fitting parameters for the two sets of radial velocities.

\begin{table}
\centering
\caption[]{Orbital solutions.}
\label{tabrvfit}
\begin{tabular}{lcc}
\hline
\hline
Parameter                  & H~{\sc i}+He~{\sc i}+He~{\sc ii}& He~{\sc i}+He~{\sc ii} (adopted) \\
\hline
e                          &        0.63$\pm$0.11            &        0.72$\pm$0.15         \\
$\omega$ (deg)             &         7.8$\pm$13.9            &        21.0$\pm$12.7         \\
$\gamma$ (km s$^{-1}$)     &     $-$48.4$\pm$1.8             &     $-$40.2$\pm$1.9          \\
$\phi_{\rm peri}$          &        0.22$\pm$0.02            &        0.23$\pm$0.02         \\
$K_1$ (km s$^{-1}$)        &        18.1$\pm$3.3             &        22.6$\pm$6.3          \\
$a_1 \sin i$ (R$_{\odot}$) &         7.4$\pm$1.6             &         8.2$\pm$2.9          \\
$f(M)$ (M$_{\odot}$)       & 0.0076$^{+0.0060}_{-0.0038}$    & 0.0107$^{+0.0163}_{-0.0077}$ \\
$\sigma$ (km s$^{-1}$)     &           7.8                   &            8.5               \\
\hline
\end{tabular}
\end{table}

\begin{figure}
\centering
\includegraphics[width=84mm]{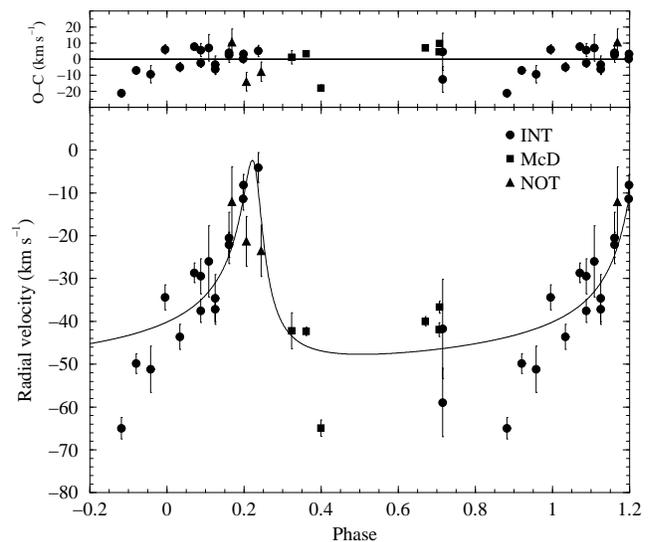}
\caption{Radial velocity curve obtained by cross-correlating with $\tau$ Sco
using the He~{\sc i} and He~{\sc ii} lines in our spectral range. The
best-fitting solution, using an eccentric orbit, is overplotted. The upper panel
shows the residuals of the fit.}
\label{figrv}
\end{figure}

\begin{figure}
\centering
\includegraphics[width=84mm]{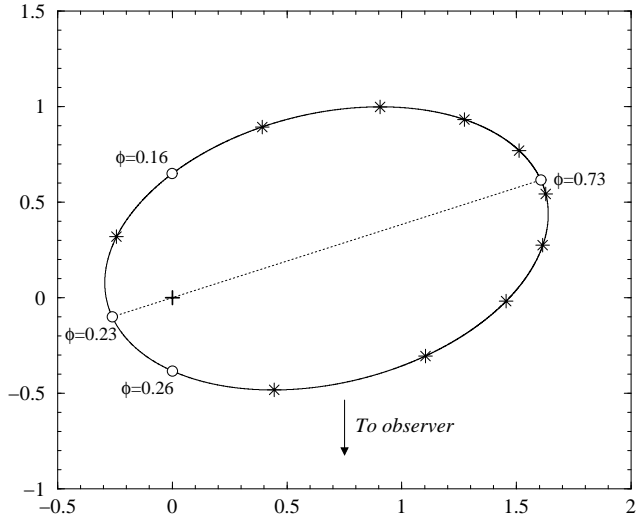}
\caption{Relative orbit of the compact object around the optical component,
which lies in the ellipse focus at (0,0). Since the actual scale of the orbit
is not known (both the inclination and the mass ratio cannot be determined
with the present single-lined radial velocity data), the coordinates are in
units of the orbital semi-major axis. Relevant phases such as the periastron,
apastron, and conjunctions are indicated. Stars mark 0.1-phase intervals.}
\label{figorbit}
\end{figure}

The two solutions yield orbital parameters that are consistent within
$1\sigma$ except for the systemic velocity, which is more negative when
the Balmer lines are included in the correlation. This suggests that the
Balmer lines may be contaminated by some blueshifted absorption,
consistent with the suspicion that the Be primary might have a
significant wind. Therefore, we prefer to adopt the solution that
considers the velocities derived from the He lines alone, which should be
largely free from contaminating emission. Figure \ref{figrv} shows these
velocities, together with the best-fitting orbital solution. To better
illustrate the resulting orbit, we show in Fig. \ref{figorbit} the
relative motion of the compact object around the optical companion as seen
from above (i.e., no inclination value assumed). Some relevant phases
(i.e., periastron, apastron, and conjunctions) are labelled.

\begin{figure}
\centering
\includegraphics[width=68mm,angle=-90]{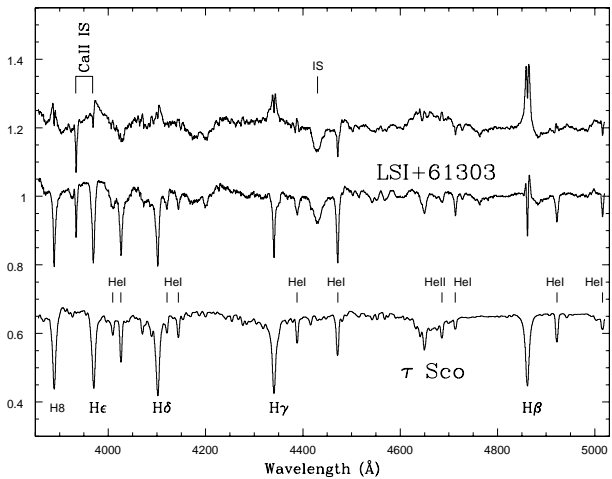}
\caption[]{Doppler corrected average of LS~I~+61~303 in the rest frame of the
optical star together with the B0~V template $\tau$ Sco and the residual after
an optimal subtraction (top). Features labeled "IS" are of insterstellar
origin. The template has been broadened by 113 km s$^{-1}$ and scaled by a
factor 0.65 to match the depth of the absorption lines in LS~I~+61~303.}
\label{figspec}
\end{figure}

We can now use our orbital solution to Doppler-shift all the individual
spectra and produce an average spectrum in the reference frame of the
primary (see Fig. \ref{figspec}). This can be compared to the spectrum of
the B0~V template $\tau$ Sco and we find that LS~I~+61~303 is best fitted
when the template is scaled by a factor 0.65. In other words, the results
indicate that the primary contributes 65 percent of the total light. The
rest can possibly be attributed to emission from the disc around the Be
star. The residual of the spectral subtraction indeed shows broad
double-peaked Balmer emission features (i.e., characteristic of a disc
geometry) with possible narrow reversal absorptions superimposed, in
accordance with observations of Be-shell stars
\citep[e.g.,][]{porriv03,slett88}.
However, inspection of high resolution H$\alpha$ profiles published in 
literature (e.g. Paredes et al. 1994) show clear double-peaked profiles 
with shallow central absorptions, reminiscent of the intermediate inclination 
Be case proposed by Hanuschik (1996). Therefore, contamination of the residual 
Balmer profiles by shell lines can almost be ruled out and we interpreted the 
observed profiles as being shaped by kinematic (doppler) broadening.

\section{Discussion}

\begin{figure}
\centering
\includegraphics[width=84mm]{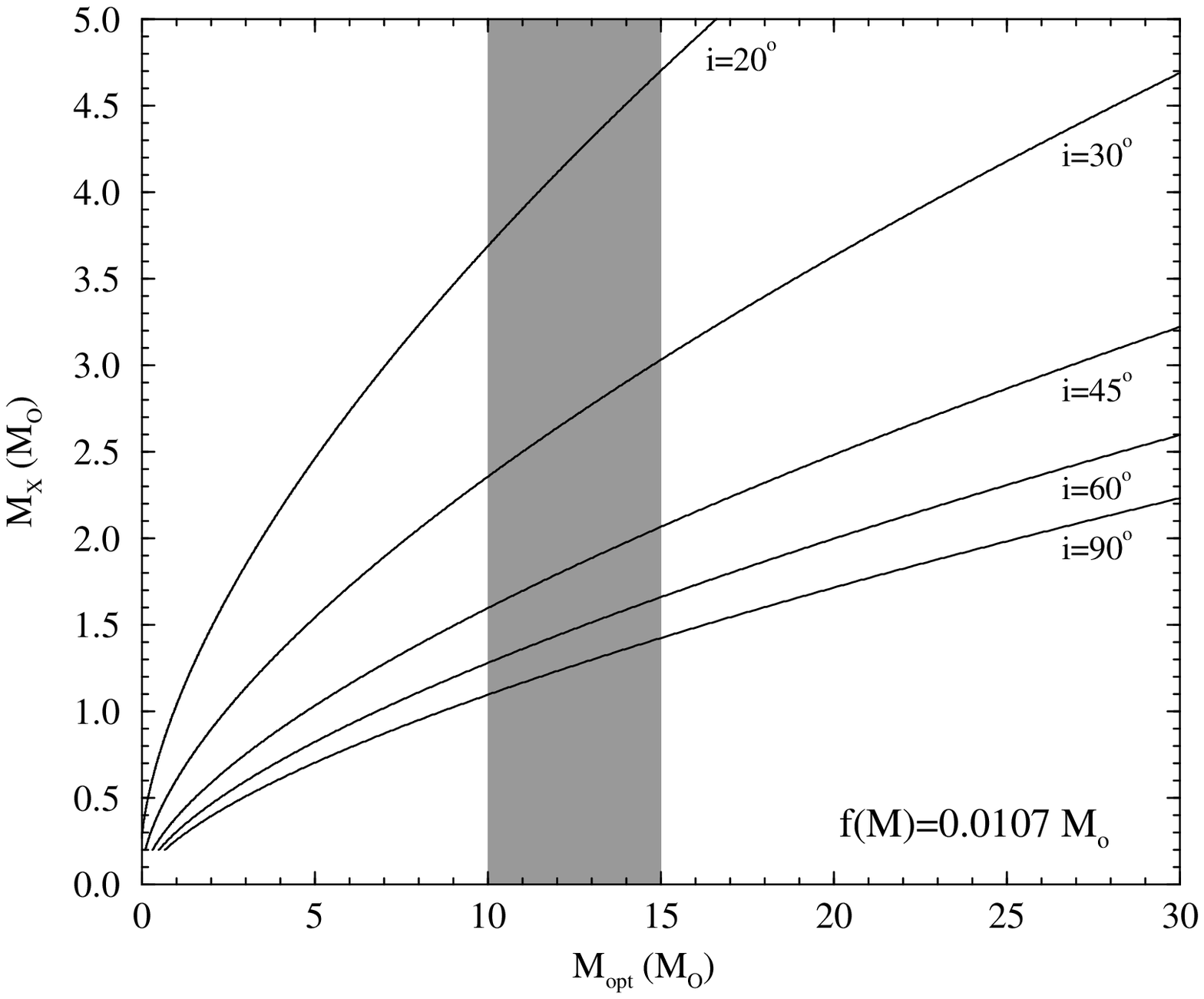}
\includegraphics[width=41.5mm]{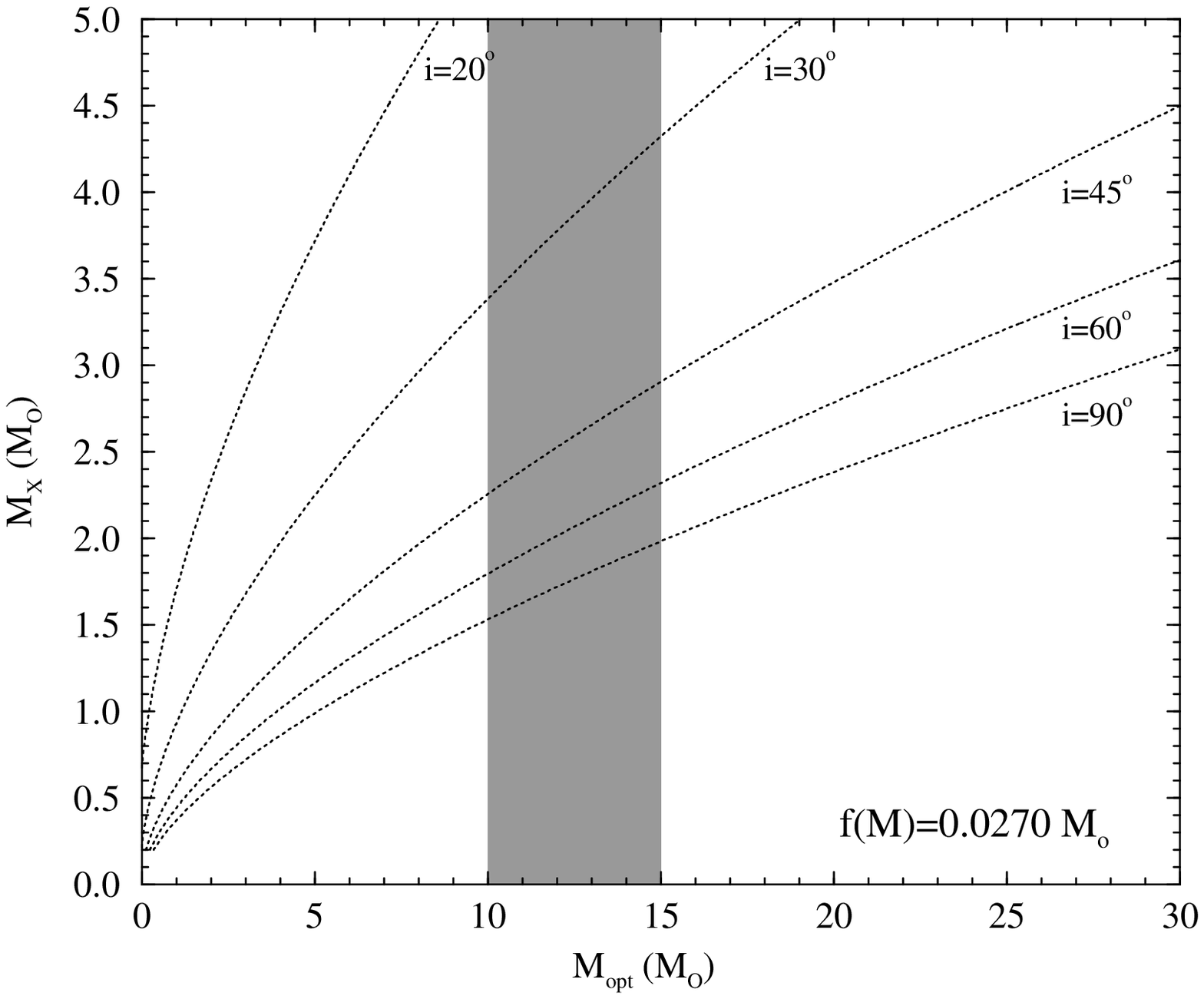}
\includegraphics[width=41.5mm]{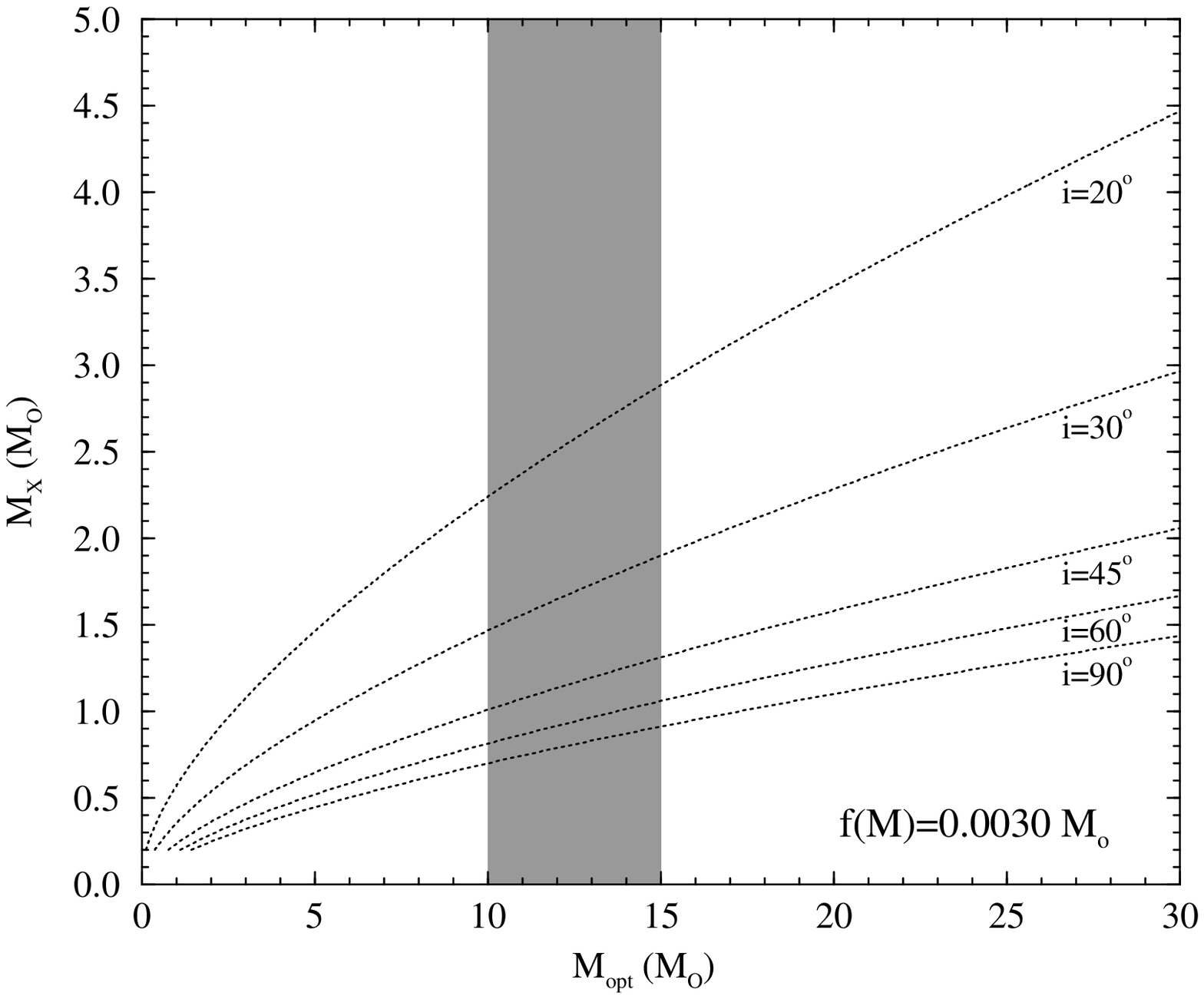}
\caption[]{Mass constraints for the two stars in LS~I~+61~303 derived from our
orbital solution. The lines have been calculated for different inclination
values and using our adopted mass function (top), plus (bottom left) and minus
(bottom right) one sigma. The shaded region represents the interval of likely
masses for the B0~V optical companion (see text).}
\label{figconst}
\end{figure}

Our radial velocity data and orbital solution yields a number of orbital
parameters but does not provide a full characterization of the system's
physical properties. For example, only the mass function and not the mass
ratio can be determined because only the velocities of the optical
component are measurable. In addition, the orbital inclination remains
unknown as usual in any radial velocity curve solution. Thus, with the
available data we can only plot Fig. \ref{figconst} that relates the mass
of the compact object ($M_{\rm X}$) and the optical companion ($M_{\rm
opt}$) for several values of the orbital inclination. Note that the plot
has been computed with our adopted radial velocity solution that yields
$f(M)$$=$$0.0107^{+0.0163}_{-0.0077}$~M$_{\odot}$. To further constrain
the system's properties we must employ additional external information.

The expected masses of the components can be used to constrain the
system's geometry. For example, the mass of the optical component of the
system can be inferred from its atmospheric properties. Discussions on the
most likely mass values for the primary star can be found in
\citet{Hutchings&crampton81}, \citet{Marti95}, \citet{Punsly99}, and
\citet{Massi04}, overall favouring a mass interval of $M_{\rm
opt}=$10--15~M$_{\odot}$ (shaded area in Fig. \ref{figconst}). With this
constraint, the compact object would be a neutron star for inclinations
$25^{\circ}$$\la$$i$$\la$$60^{\circ}$ and a black hole if
$i$$\la$25$^{\circ}$.

There are several options to obtain independent information on the
inclination. For example, we have measured $v \sin i \simeq 113$ km
s$^{-1}$, which is about 3 times smaller than the value reported by
\cite{Hutchings&crampton81}. We note that our instrumental resolution is
higher than that of \citeauthor{Hutchings&crampton81} (83 km~s$^{-1}$
versus $\sim$130 km~s$^{-1}$), which would favour our determination. This
is relevant for the orbital inclination (assuming that the Be star's spin
axis and the orbital axis are aligned) because 
current observational constraints indicate that  Be stars 
rotate at $\simeq$0.7--0.8 times\footnote{With a likelihood that this might 
be underestimated due to equatorial gravity darkening effects 
(see Townsend, Owocki \& Howarth 2004, Porter \& Rivinius 2003).} 
the critical rotational velocity\citep{por96}.   
In the case of the optical component of LS~I~+61~303 this
would be $v_{\rm crit}\sim$500--550 km~s$^{-1}$. Therefore, if the primary
is a normal Be star, we are viewing the binary at a low inclination angle
of $\sim$15--20$^{\circ}$ 
but we cannot rule out that the Be star may be rotating at 
substantially lower speed than $v_{\rm crit}$. 
However, the lack of clear shell lines in our spectra, such as 
FeII $\lambda$4584, points to a rough upper limit $i \lesssim 60^{\circ}$. 
A lower limit to the 
inclination of $i\sim$10--15$^{\circ}$ is obtained from the restriction
that the rotational velocity of the optical companion should not exceed
0.9 times the critical value. Perhaps future accurate modelling of the
spectral line profiles could provide a better constraint on the viewing
angle of the Be star and thus on the orbital inclination.

Additional information on the inclination might come from the geometry of
the relativistic jet as derived from the VLBI images of \cite{Massi04} by
assuming perpendicularity between the ejection direction and the orbital
plane. However, because of the possible precession reported by these
authors, values in the full range $i=$0--90$^{\circ}$ could be feasible.

To conclude with the discussion of the orbital inclination, it seems that
a crude value of $\sim$30$^{\circ}$ (with an uncertainty of some
20$^{\circ}$) is the best compromise given the available constraints at
this point.  It is clear that further observations will be required to
better define this important parameter of the system. With the current
knowledge, both a neutron star and black hole are equally plausible as the
compact object in the LS~I~+61~303 system.

But there is one very relevant parameter that is indeed well-constrained
by our orbital solution, and that is the phase of the periastron passage.
We obtain a value of $\phi_{\rm peri}=0.23\pm0.02$, in good agreement with
that obtained by \cite{Hutchings&crampton81} ($\phi_{\rm peri}=0.2$), when
the phase origin is set by the radio ephemeris of \cite{Gregory02b}. A key
question is how this periastron passage phase relates to the emission
variability of LS~I~+61~303. There have been a number of studies of the
occurrences of radio outbusts in LS~I~+61~303 and there is general
consensus in that the radio emission peaks are distributed over a wide
phase interval of about 0.45--0.95 (e.g., \citeauthor*{Paredes90}
\citeyear{Paredes90}; \citeauthor{Gregory02} \citeyear{Gregory02}). In
contrast, X-ray maxima seem to occur at somewhat earlier phases of
0.43--0.47, although this is based on only two observations. From three
well-studied radio outbursts, \cite{Gregory02b} was able to determine the
phases at which the onset of the outbursts occurred and obtained values in
the interval 0.33--0.40. With our new determination of the periastron
phase, it is now clear that a delay exists beween the closest approach of
the system's components and the radio and X-ray outbursts. The onset of
the outbursts appears to take place about 2.5--4 days after periastron
passage. The delay is likely due to inverse Compton energy losses of
relativistic electrons in the radio jets in the periastron vicinity (see
\citeauthor{Massi04b} \citeyear{Massi04b} for details). This now
well-established result has strong implications on the models devised to
explain the emission behaviour of LS~I~+61~303.

\section{Acknowledgments}

We thank the anonymous referee for helpful comments to the manuscript. 
J.~C. acknowledges support from the Spanish MCYT grant AYA2002-0036.  
I.~R. acknowledges support from the Spanish Ministerio de Ciencia y
Tecnolog\'{\i}a through a Ram\'on y Cajal fellowship. J.~M.~P. and J.~M.
acknowledge partial support by the DGI of the Ministerio de Ciencia y
Tecnolog\'{\i}a (Spain) under grant AYA2001-3092, as well as partial
support from the European Regional Development Fund (ERDF/FEDER). J.~M. is
also supported by the Junta de Andaluc\'{\i}a (Spain) under project
FQM322. MOLLY and DOPPLER software developed by T. R. Marsh is gratefully
acknowledged. The INT is operated on the island of La Palma by the Royal
Greenwich Observatory in the Spanish Observatorio del Roque de Los
Muchachos of the Instituto de Astrof\'\i{}sica de Canarias (IAC). The NOT
is operated on the island of La Palma jointly by Denmark, Finland,
Iceland, Norway and Sweden in the Spanish Observatorio del Roque de Los
Muchachos of the IAC.

\end{document}